# Entanglement enhancement of two different magnon modes via nonlinear effect in cavity magnomechanics

Ke Di,[1] Xi Wang,[1] Shuai Tan,[1] Yinxue Zhao,[2] Yu Liu,[1] Anyu Cheng[1] and Jiajia Du[1,*]

[1]*Chongqing University of Post and Telecommunications, Chongqing, 400065, China*
[2]*Wuhan Social Work Polytechnic, Wuhan, 430079, China*
*\*author dujj@cqupt.edu.cn*

We present a scheme to enhance two different magnon modes entanglement in cavity magnomechanics via nonlinear effect. The scheme demonstrated that nonlinear effects enhance entanglement of the two magnon modes. Moreover, the entanglement of the two magnon modes is also significantly enhanced by microwave parametric amplification (PA) and magnon self-Kerr nonlinearity. Not only dose nonlinear effect enhances the strength of entanglement, but it also increases the robustness of entanglement against temperature. Our proposed scheme plays an important role in the research of fundamental theories of quantum physics and quantum information processing theory.

The cavity-magnon system has garnered significant interest in the field of quantum information and has been the subject of growing investigation in recent years [1,2]. Yttrium iron garnet (YIG) plays a significant role in the cavity-magnon system, serving as a crucial component for investigating the profound interactions between light and matter on a distinct platform. The successful achievement of coupling between Kittel modes and microwave cavity photons in YIG spheres can be attributed to their notable characteristics, including high spin density and low dissipation rate [3]. This coupling facilitates the transmission of coherent information between distinct information carriers. The interaction between the coherent photon-spin ensemble in the YIG crystal is significantly amplified due to its high spin density, and it can potentially reach the ultrastrong coupling regime [4–11]. Numerous interesting phenomena have been explored between the magnon and the microwave cavity, such as the observation of magnon gradient memory to store information in the magnon dark mode [12], cavity spintronic [13,14], cooperative polariton dynamic [15] and the exceptional point [16,17].

In recent times, a number of suggestions have emerged regarding the generation of entangled states involving two magnon modes within ferrimagnetic YIG spheres, employing various processes. The entanglement of magnon modes is a highly significant resource that is utilized in several applications within the field of quantum information science. These applications include quantum logical operations, quantum metrology, and the fundamental principles of quantum mechanics [18-21]. The magnon mode can undergo squeezing when the cavity is driven by a squeezed vacuum microwave field. This activation of magnetostrictive interactions occurs when red detuned microwave fields drive the magnon mode. Consequently, the squeezing effect can be translated to mechanical modes [22]. The placement of a parametric amplifier within the hybrid cavity serves to enhance the squeezing of the cavity field [23]. The focal point of this current study is the primary subject matter. In a recent study, Li et al. [24-25] examined the phenomenon of entanglement in the context of two magnon modes and two vibrational modes of YIG spheres.

The subsequent sections of the paper are structured in the following manner. In Section 2, we provide a cavity magnomechanical system consisting of two distinct states of YIG spheres. We provide comprehensive guidelines on solving the quantum Langevin equation (QLE) and computing entanglement using Hamiltonians. In Section 3, a linearized approach is employed to examine the dynamics of the system. The primary findings on the entanglement between the two magnon modes in various phases are presented. In Section 4, the ideal squeezing parameters are selected in order to enhance the entanglement between two magnon modes. In conclusion, we arrive at our last remarks in Section 5.

We have established a hybrid cavity magnomechanical system consisting of a microwave cavity mode, two magnon modes, a mechanical mode, and a PA, as seen in figure 1. In hybrid cavity magnomechanical system, microwave cavity mode can be strongly coupled through parameter amplifier. The collective motion of a large number of spins reflects the magnon mode in macroscopic ferromagnets, coupled to a single microwave cavity mode. Such a system of two YIG spheres (without involving the mechanical mode) has been used to study magnon dark modes [3] and high-order exceptional points [8]. The magnetic dipole interaction mediates the coupling between magnons and cavity photons, and this coupling can be extremely strong [26–29]. The vibration of the YIG sphere caused by magnetostrictive force, which leads to geometric deformation of the sphere and the generation of phonon modes. In general, the magnetostrictive interaction is of different types depending on the resonance frequencies of the magnon and phonon modes [32]. Due to the mechanical mode frequency being much lower than the magnon mode frequency, the magnon-phonon coupling is very small [29], but it can be significantly enhanced through a strong microwave field [30,31]. The magnomechanical coupling strength is sensitive to the direction of the bias magnetic field [29]. We adjust the direction of the two bias magnetic fields without changing the amplitude of the magnetic field (see figure 1). YIG 2 will not generate phonons when selecting the appropriate bias magnetic field direction.

The Hamiltonian of the system under rotating-wave approximation in a frame rotating with the frequency of the drive field is given by

$$H_0 = \hbar\omega_c c^\dagger c + \sum_{j=1,2} \hbar\omega_{mj} m_j^\dagger m_j + \frac{\hbar\omega_b}{2}(q^2 + p^2)$$
$$+ K m^\dagger m m^\dagger m + \sum_{j=1,2} g_j(cm_j^\dagger + c^\dagger m_j) + G_0 m_1^\dagger m_1 q$$
$$+ i\Omega(m_1^\dagger e^{-i\omega_0 t} - m_1 e^{i\omega_0 t}) + iG(e^{i\theta}c^{\dagger 2} - e^{-i\theta}c^2) \quad (1)$$

Where c and $c^\dagger$ ($m_j$ and $m_j^\dagger$), [O, $O^\dagger$]=1, O = c ($m_j$). $\omega_c$, $\omega_{m_j}$, and $\omega_b$ are the resonance frequency of the cavity, magnon, and mechanical modes. The first (second) term describes the energy of the cavity mode (magnon modes). The magnon frequency $\omega_{mj} = \gamma H_j$ is determined by the bias magnetic field $H_j$ where $\gamma/2\pi = 28$ GHz/T is the gyromagnetic ratio. The third term denotes the energy of two mechanical vibration modes, and $q_j$ and $p_j$ ($[q_j, p_j] = i$) are the dimensionless position and momentum of the vibration mode j, modeled as a mechanical oscillator. The fourth term represents the self-Kerr of the magnon mode, and K is the self-Kerr coefficient. The coupling rate $g_j$ denotes the linear coupling between the cavity and the jth magnon mode, and $G_0$ represents the single-magnon magnomechanical coupling rate. The Rabi frequency $\Omega = \frac{\sqrt{5}}{4}\gamma\sqrt{N}B_0$ [32] denotes the coupling strength, $B_0$ is the amplitude of drive magnetic field with the first magnon mode. N = ρv describes the total number of spins, and ρ=4.22×10²⁷ m⁻³ is the spin density of the YIG, and V is the volume of the sphere. The last term denotes the nonlinear gain of the PA, and G is the parameter with θ as the phase of the driving field.

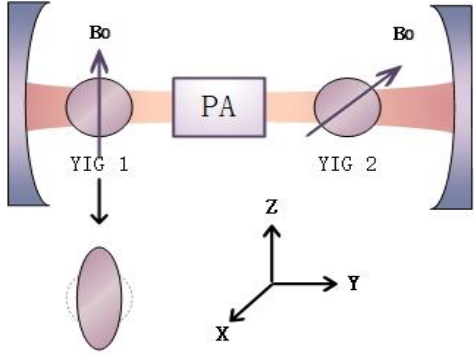

FIG. 1. Schematic of a single-mode cavity consists of the PA and two YIG spheres. The two YIG pheres are placed near the maximum magnetic field of the cavity mode in the cavity. Two magnon modes are excited in the spheres coupled to the cavity mode. The deformation of the YIG sphere is controlled by adjusting the direction of the bias magnetic field. The magnetic field of cavity mode along the y-direction of the coordinate axis. External bias magnetic field along the coordinate axis z-direction

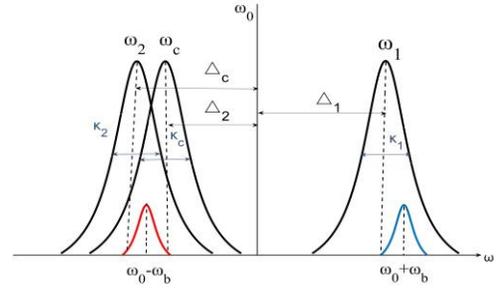

FIG.2. The frequency and linewidth of the system are shown in the figure. The mechanical motion of frequency $\omega_b$ scatters photons onto the two sidebands at $\omega_0 \pm \omega_b$. The magnon mode $m_1$ is resonant with the blue (anti-Stokes) sideband, and both the cavity and the magnon mode $m_2$ with frequency $\omega_2$ are resonant with the red (Stokes) sideband. The two magnon modes are prepared in an entangled state through sideband resonance.

The QLEs of the system are given by

$$\dot{c} = -(i\Delta_c + \kappa_c)c - i\sum_{j=1,2}g_j m_j + \sqrt{2\kappa_c}c^{in} + 2Ge^{i\theta}c^\dagger$$
$$\dot{m}_1 = -(i\Delta_1 + \kappa_1)m_1 - iG_0 m_1 q + \Omega - 2iKm^\dagger mm + \sqrt{2\kappa_1}m_1^{in}$$
$$\dot{m}_2 = -(i\Delta_2 + \kappa_2)m_2 - ig_2 c + \sqrt{2\kappa_2}m_2^{in}$$
$$\dot{q} = \omega_b p$$
$$\dot{p} = \omega_b q - \gamma_b p - G_0 m_1^\dagger m_1 + \xi \quad (2)$$

$\kappa_c$, $\kappa_{j1}$, and $\gamma_b$ are the dissipation rates of the cavity, magnon, mechanical modes, respectively, and $c^{in}$, $m_j^{in}$ are input noise operators affecting the cavity and magnon modes, respectively, which are zero mean and characterized by the following correlation functions. The Langevin force operator ξ, which accounts for the Brownian motion of the mechanical mode, is autocorrelated as $\langle \xi(t)\xi(t') + \xi(t')\xi(t)\rangle/2 = \gamma_b[2N_b(\omega_b) + 1]\delta(t - t')$, where we have made the Markov approximation, which is a good approximation for a mechanical oscillator of a large quality factor $Q_b = \omega_b/\gamma_b \gg 1$ [34]. $G_{mb} = i\sqrt{2}G_0\langle m_1\rangle$ is the effective magnomechanical coupling rate.

This allows us to linearize the dynamics of the system around the steady-state values by writing any operator as $O = \langle O\rangle + \delta O$, (O = a, $m_j$, q, p), and neglecting small second-order fluctuation terms. Since we are particularly interested in the quantum correlation properties of the two magnon modes, we focus on the dynamics of the quantum fluctuations of the system. The linearized QLEs describing the fluctuations of the system quadratures $(\delta X, \delta Y, \delta x_1, \delta y_1, \delta x_2, \delta y_2, \delta q, \delta p)$, with $\delta X = (\delta c + \delta c^\dagger)/\sqrt{2}$, $\delta Y = i(\delta c^\dagger - \delta c)/\sqrt{2}$, $\delta x_j = (\delta m_j + \delta m_j^\dagger)/\sqrt{2}$ and $\delta y_j = (\delta m_j^\dagger - \delta m_j)/\sqrt{2}$, can be written in the form of

$$\dot{u}(t) = Au(t) + n(t) \quad (3)$$

where $u(t) = [\delta X, \delta Y, \delta x_1, \delta y_1, \delta x_2, \delta y_2, \delta q, \delta p]^T$, $n(t)$ is the vector of input noises, and the drift matrix A is given by

$$A = \begin{pmatrix} -\kappa_c + 2G\cos\theta & \Delta_c + 2G\sin\theta & 0 & g_1 & 0 & g_2 & 0 & 0 \\ -\Delta_c + 2G\sin\theta & -\kappa_c - 2G\cos\theta & -g_1 & 0 & -g_2 & 0 & 0 & 0 \\ 0 & g_1 & -\kappa_1 & \Delta_1 - k & 0 & 0 & -G_{mb} & 0 \\ -g_1 & 0 & -\Delta_1 - k & -\kappa_1 & 0 & 0 & 0 & 0 \\ 0 & g_2 & 0 & 0 & -\kappa_2 & \Delta_2 & 0 & 0 \\ -g_2 & 0 & 0 & 0 & -\Delta_2 & -\kappa_2 & 0 & 0 \\ 0 & 0 & 0 & 0 & 0 & 0 & 0 & \omega_b \\ 0 & 0 & 0 & G_{mb} & 0 & 0 & -\omega_b & -\gamma_b \end{pmatrix}$$
(4)

The steady state of the quantum fluctuations of the system is thus a continuous variable four-mode Gaussian state, which is completely characterized by an 8×8 covariance matrix (CM) which is defined as $V_{ij} = \langle u_i(t)u_j(t') + u_j(t')u_i(t)\rangle/2$ (i,j = 1,2,…,8). The stationary CM can be straightforwardly obtained by solving the Lyapunov equation [35,36].

$$AV + VA^T = -D \quad (5)$$

The diffusion matrix D is given by
$$D = diag[\kappa_c(2N_c + 1), \kappa_c(2N_c + 1), \kappa_1(2N_1 + 1),$$
$$\kappa_1(2N_1 + 1), \kappa_2(2N_2 + 1), \kappa_2(2N_2 + 1), 0,$$
$$\gamma_b(2N_b + 1)] \quad (6)$$
which is defined by $D_{ij}\delta(t-t') = \langle n_i(t) n_j(t') + n_j(t) n_i(t')\rangle/2$). We adopt the logarithmic negativity [37] to quantify the magnon entanglement, The logarithmic negativity is defined as [38]
$$E_N \equiv max[0, -ln2\widetilde{v_-}] \quad (7)$$
where $\widetilde{v_-} = 2^{-1/2}\{\Sigma(V) - [\Sigma(V)^2 - 4detV_0]^{1/2}\}^{1/2}$, $V_0$ is the $4 \times 4$ CM associated with the two magnon modes mode. $V_0 = [V_1, V_{12}; V_{12}, V_2]$, with $V_1, V_{12}$ and $V_2$ being the $2\times 2$ blocks of $V_0$, and $\Sigma V \equiv detV_1 + detV_2 - 2detV_{12}$.

In this study, we describe the outcomes of numerical simulations conducted on entanglement phenomena. Our findings reveal that the inclusion of PA not only amplifies entanglement at resonance positions but also induces entanglement throughout a broader spectrum of parameter spaces. Figure 3 illustrates the correlation between the entanglement of the two magneton modes and some crucial parameters of the system.

We adopt the experimentally feasible parameters:[28] $\omega_c/2\pi = 12\ GHz, \omega_b/2\pi = 10\ MHz, \gamma_b/2\pi = 10^2\ Hz, \kappa_c/2\pi = 1\ MHz, \kappa_1/2\pi = \kappa_2/2\pi = 1\ MHz$, $g_1/2\pi = 3.2\ MHz$, $G_{mb}/2\pi = 4.8\ MHz, G/2\pi = 10^6$, and at a low temperature T=10mk. We have $g_{1,}^2 g_2^2 \ll |\widetilde{\Delta}_1 \Delta_c|, |\Delta_2 \Delta_c| \simeq \omega_b^2$. When the drive magnetic field $B_0 \simeq 3.9 \times 10^{-5}$ T for $G_0/2\pi = 0.3Hz$ and drive power p ≈ 8.9 mW, the effective magnomechanical coupling rate $G_{mb}/2\pi = 4.8MHz$.

The initial magnon mode is primarily responsible for inducing a substantial cooling effect on the mechanical mode. This cooling effect arises due to the presence of quantum entanglement, which is observed exclusively when the mode is occupied by low-energy thermal phonons. The phenomenon of magnomechanical coupling, denoted as $G_{mb}$, gives rise to the entanglement between magnons and phonons.

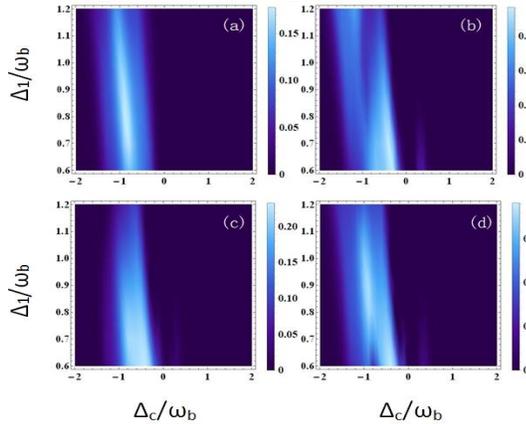

FIG. 3. Density plot of the entanglement $E_{m1m2}$ between two magnon modes. dimensionless detunings $\Delta_c/\omega_b$ and $\Delta_2/\omega_b$ (a) G = 0, (b)–(d) $G/2\pi = 10^6$. The phase θ is (b) 0, (c) π/2, and (d) π, respectively. We take $\Delta_2 = \Delta_c$ and $g_2/2\pi = 2.6$ MHz.

This entanglement is subsequently transferred to the cavity-magnon system, resulting in the entanglement between the cavity mode and the magnon mode, denoted as $m_1$ [30]. Three distinct values of θ have been selected in order to analyze the impact of varying phase values on the entanglement, as documented in reference [23].

We present the entanglement $E_{m1m2}$ between two magnon modes as a function of dimensionless detunings $\Delta_c/\omega_b$ and $\Delta_1/\omega_b$ in. This is confirmed by Fig. 3 which shows that the best case is the magnon mode with cavity resonance, $\Delta_2 \simeq \Delta_c \simeq \omega_b$. Fig. 3(a) show $E_{m1m2}$ without the PA. The enhancement is obtained for θ = π/2 in the presence of PA as shown in Fig. 3(c), and θ = π in the presence of PA as shown in Fig. 3(d). The enhancement in entanglement is because parametric gain G and phase θ obtain squeezed vacuum microwave field. The spatial extent of the presence of entanglement is also significantly increased. The eigenvalues of the drift matrix A are affected by the gain G and the phase θ, which also affect the steady-state solution of the cavity mode. There`fore, the value of the phase θ associated with the PA driving field also plays a considerable role in the dynamics of the system [27].

The phase value may lead to maximum noise rejection, resulting in maximum entanglement, or it may lead to noise amplification, resulting in reduced entanglement [see Figs. 3(b)]. The entanglement of all three choices of the PA phase with maximum enhancement at θ = π/2. We obtain a maximum entanglement value of $E_{m1m2} \simeq 0.19$ from $E_{m1m2} \simeq 0.23$ from Fig. 3(c), which is comparable to the non-PA entanglement value of $E_{m1m2} \approx 0.17$.

The robustness of magnon entanglement against ambient temperature is demonstrated in Figure 4, where it is observed to persist up to around 200 mK.

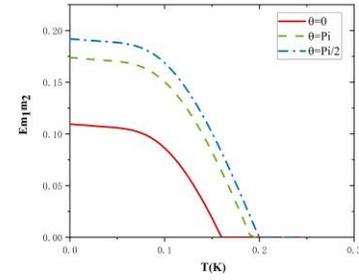

FIG. 4. Magnon entanglement $E_{m1m2}$ temperature for the three cases of θ = 0 (black dashed line), θ = π/2 (blue dashed line) and θ = π (red dashed line). We take an optimal detuning $\Delta_c = -0.9\omega_b$.

In our investigation, we provide another non-linear solution to enhance the entanglement of continuous variables, achieved by compressing the magnon mode. Due to the existence of Kerr nonlinear term $Km^\dagger m m^\dagger m$, which bring the nonlinear effect in that case of strong magnon drive. By selecting the optimal compression parameters, the entanglement value of two different magnon modes is increased [39-40].

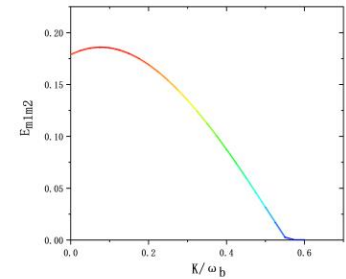

FIG. 5. Entanglement of two magnon modes versus squeezing parameter $K/\omega_b$. We adopt the optimal detuning parameters $\Delta_a = -0.9\omega_b, \Delta_{m1} = 0.85\omega_b, \Delta_{m2} = -0.9\omega_b$.

The curve of entanglement value changing with squeezing parameters is a quadratic function, This demonstrates the nonlinearity of the Kerr effect. With the increase of squeezing parameters, the change of entanglement value is shown in Figure 5. In the red curve section, the squeezing of the magnon mode ($m_1$) enhances the entanglement Em1m2, as shown in Figure 5. When $K/\omega_b > 0.6$ the entanglement value decreases to zero.

In conclusion, we investigate a scheme to enhance the entanglement of two magnon modes in cavity magnomechanics via nonlinear effects. Compared to the Kerr effect, PA significantly enhances the entanglement of two magnetic magnon modes in selecting the appropriate phase. The robustness of the entanglement of the two magnon modes against temperature is also enhanced in the optimal phase. PA is capable of generating squeezed microwave field to enhance entanglement, and Kerr effect enhances entanglement through magnon squeezing. There were able to be verified experimentally. Our work will play a key role in quantum information processing in hybrid quantum systems.

**Funding.** National Natural Science Foundation of China (Grant Nos. 11704053, 52175531); the National Key Research and Development Program of China (Grant No. 2021YFC2203601).

**Disclosures.** The authors declare no conflicts of interest.

**Data availability.** The data that support the findings of this study are available within the article.